\documentclass[aps,prl,nopacs,superscriptaddress,twocolumn,floatfix]{revtex4}
\usepackage{graphicx}% Include figure files
\usepackage{dcolumn}% Align table columns on decimal point
\usepackage{bm}% bold math
\usepackage{color}
\usepackage[breaklinks=true,colorlinks,citecolor=blue,linkcolor=blue,urlcolor=blue]{hyperref}
\begin{document}
\title{Artificial graphene as a tunable Dirac material}
\author{Marco Polini}
\email{m.polini@sns.it}
\affiliation{NEST, Istituto Nanoscienze-CNR and Scuola Normale Superiore, I-56126 Pisa, Italy}
\author{Francisco Guinea}
\affiliation{Instituto de Ciencia de Materiales de Madrid (CSIC), Sor Juana In\'es de la Cruz 3, E-28049 Madrid, Spain}
\author{Maciej Lewenstein}
\affiliation{ICFO - Institut de Ci\`encies Fot\`oniques, Mediterranean Technology Park, Av. Carl Friedrich Gauss 3, E-08860 Castelldefels, Barcelona, Spain}
\affiliation{ICREA - Instituci\'o Catalana de Recerca i Estudis Avan\c cats, 08010 Barcelona, Spain}
\author{Hari C. Manoharan}
\affiliation{Department of Physics, Stanford University, Stanford, California 94305, USA} 
\affiliation{Stanford Institute for Materials and Energy Sciences, SLAC National Accelerator Laboratory, Menlo Park, California 94025, USA}
\author{Vittorio Pellegrini}
\affiliation{NEST, Istituto Nanoscienze-CNR and Scuola Normale Superiore, I-56126 Pisa, Italy}
\begin{abstract}
Artificial honeycomb lattices offer a tunable platform to study massless Dirac quasiparticles and their topological and correlated phases. Here we review recent progress in the design and fabrication of such synthetic structures focusing on nanopatterning of two-dimensional electron gases in semiconductors, molecule-by-molecule assembly by scanning probe methods, and optical trapping of ultracold atoms in crystals of light. We also discuss photonic crystals with Dirac cone dispersion and topologically protected edge states.  We emphasize how the interplay between single-particle band structure engineering and cooperative effects leads to spectacular manifestations in tunneling and optical spectroscopies.
\end{abstract}

\maketitle

\section{Introduction}

Graphene is boasting a profound impact in condensed-matter science~\cite{Netal05,Netal05b,Zetal05,geim07,NGPNG09} and technology~\cite{bonaccorso10,novoselov12,polini12}. It is an unusually perfect realization of a 2D semimetal displaying, in a wide range of energies, linearly dispersing conduction and valence bands, which touch at the so-called Dirac point. Charge neutrality pins the Fermi energy at the apex of the Dirac cones, which are particularly intriguing since they imply quasiparticles that behave like relativistic elementary particles with zero rest mass~\cite{katsnelsonbook}. Dirac cones characterize other materials. For instance, they describe the chiral conducting surface states that emerge on the surface of 3D topological insulators (TIs)~\cite{HK10,QZ11}. 

In the case of Dirac materials, the standard analysis of fermionic systems, based on concepts like ``effective mass" and energy gap, requires an extensive reformulation. The propagation of Dirac fermions shows unusual features~\cite{katsnelsonbook}, traceable to the presence of a new physical variable, the sublattice-pseudospin degree of freedom. Interactions between Dirac fermions~\cite{kotov_rmp_2012} are akin to those studied in quantum electrodynamics (QED) and therefore very different from interactions between Schr\"{o}dinger electrons in ordinary metals and semiconductors.  

\begin{figure*}[t]
\centering
\includegraphics[width=0.6\linewidth]{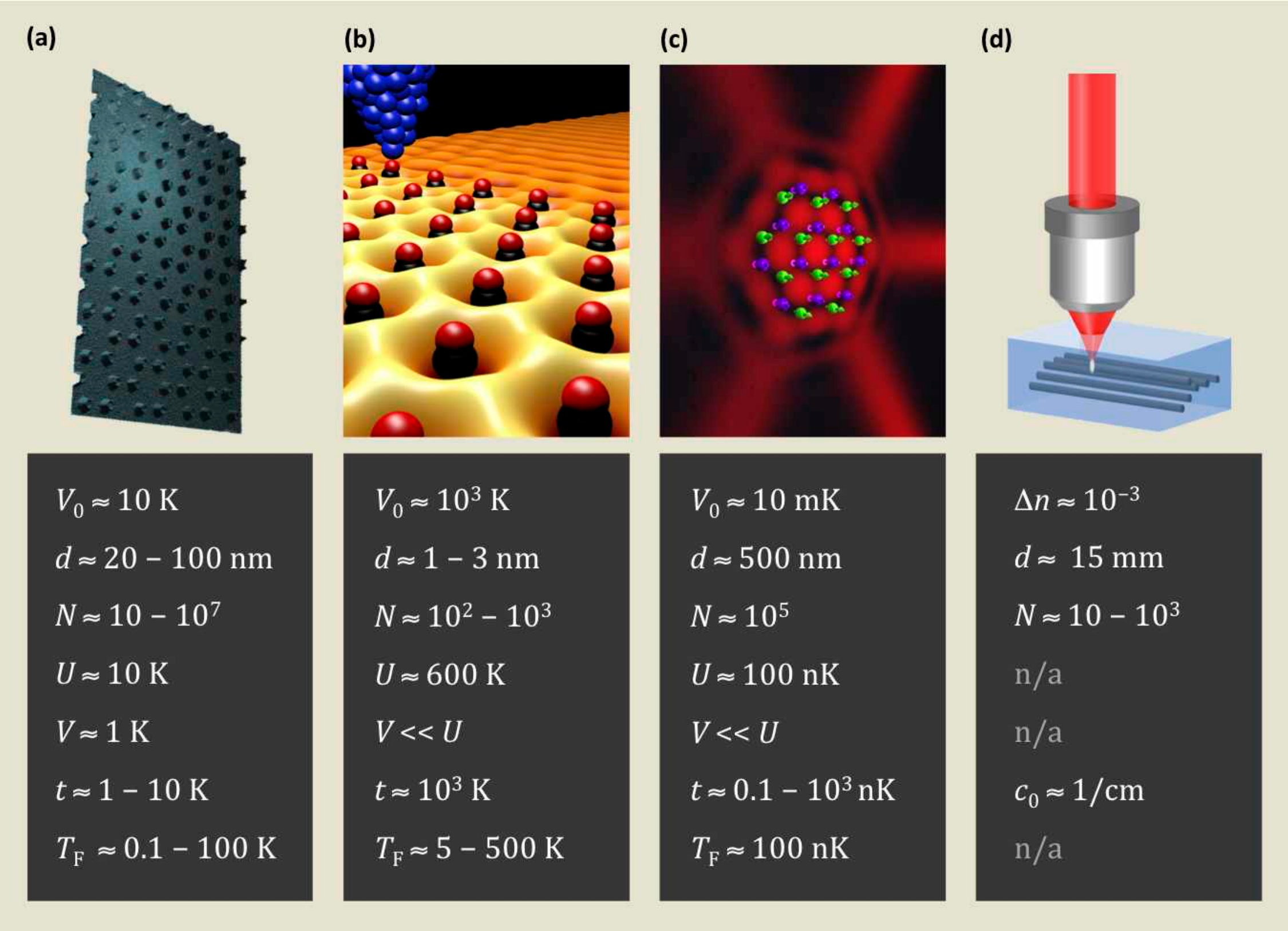}
\caption{{\bf Artificial graphene structures experimentally obtained by different methods.} 
(a) Scanning electron micrograph of the surface of a nanopatterned GaAs heterostructure. This is achieved by defining first an array of
Nickel disks with the desired geometry by e-beam nanolithography and then by etching away the material outside the disks by inductive-coupled reactive ion shallow etching and by Nickel removal. The bottom panel lists the main tunable parameters of each approach: $V_0$ is the lattice depth, $d$ is the lattice constant, and $N$ the number of sites. The parameter $U$ ($V$) is the strength of the on-site (nearest-neighbor interaction) repulsion while $t$ is the hopping energy scale. Finally, $T_{\rm F}$ is the Fermi temperature. Panels (b) and (c) refer to molecular graphene systems and optical lattices for cold atoms, respectively. In panel (b) red (black) spheres represent the Oxygen (Carbon) atoms of CO molecules while the yellow/orange surfaces represent the electron density in a honeycomb pattern. Panel (d) describes photonic honeycomb crystals induced by optical induction methods. $\Delta n$  is the change of refractive index induced by laser irradiation. The parameter $c_0$ is the coupling constant between waveguides, which plays the role of the hopping parameter $t$. There is no Hubbard gap or band gap in this system (only the one that is induced by the strain in the form of Landau levels). Panel (c) courtesy of L. Fallani.\label{figureone}}
\end{figure*}

Dirac conical singularities can actually emerge in any 2D lattice. Symmetry arguments, in fact, show that they should appear regularly at the corners of the Brillouin zone (BZ) in lattices with triangular symmetry. Therefore Dirac fermion physics and its technological exploitation should not be considered restricted to the realm of ``natural" materials---such as graphene and TIs---but could also be explored, in principle, in artificial structures displaying triangular symmetry.  Advantages of such ``artificial graphene" (AG) systems are likely to be their tunable properties---including lattice constants, hopping energies, and inter-particle interaction strength---and the possibility to design and realize artificial defects. Additionally, the creation of such structures is particularly attractive for the study of quantum phases driven by correlation effects, which in natural (single-layer) graphene seem to emerge only in ultra-high magnetic fields~\cite{kotov_rmp_2012}. 

Recent advances have demonstrated the possibility of creating AG in diverse subfields of low-energy physics. 
Current methods of design and synthesis include i) nanopatterning of ultra-high-mobility 2D electron gases (EGs)~\cite{park_nanoletters_2009,gibertini_prb_2009,desimoni_apl_2010,singha_science_2011,nadvornik_njp_2012,goswami_prb_2012}, ii) molecule-by-molecule assembly on metal surfaces by scanning probe methods~\cite{gomes_nature_2012}, iii) trapping ultracold fermionic and bosonic atoms in honeycomb optical lattices~\cite{wunsch_njp_2008,SoltanPanahi11a,Tarruell12}, and iv) confining photons in honeycomb photonic crystals~\cite{sepkhanov07,Peleg07,sepkhanov08,haldane_prl_2008}. 
Figure~\ref{figureone} summarizes these four different approaches 
and the relevant tunable parameters. Here we review the state of the art of this emerging multidisciplinary field 
and offer perspectives on possible future developments.

\section{Designing and probing Dirac bands in artificial lattices}

AG structures may not immediately compete with graphene for technological applications. However, they offer a playground to observe and comprehend physical phenomena related to Dirac energy-momentum dispersion relations in regimes that are difficult to achieve in natural graphene. In the following we discuss the most relevant AG lattices explored so far in which electrons, photons, cold atoms, and ions are the confined particles. As we elaborate below, these systems have complementary physical properties that enable the investigation of a wide range of phenomena.

\subsection{Confining electrons}

In 1970 Esaki and Tsu realized the possibility to engineer energy bands by artificially modulating the potential in one direction~\cite{esaki70}. They came up with the idea of using a semiconductor superlattice and predicted the onset of negative differential conductivity. This pioneering work stimulated a large effort worldwide focused on band-gap engineering in semiconductor heterostructures that, thanks to the refinement of nanofabrication techniques, eventually led to the development of {\it lateral superlattices}---semiconductor systems characterized by a 2D periodic potential modulation~\cite{ferry87}. 

The advent of these artificial crystals with a tunable band-structure has greatly influenced the field of 2DEGs in modulation-doped semiconductor heterostructures. Works in this area started in the late eighties and enabled the observation of Weiss oscillations~\cite{weiss}, novel conductance resonances due to quantization of the electron orbits in the 2D pattern in a magnetic field~\cite{albrecht99}, chaotic dynamics, and, more recently, led to studies of Hofstadter butterfly phenomena~\cite{albrecht01,melinte04}.  Recently, similar effects have been observed in natural graphene where the 2D periodic potential (with periodicity of the order of $\sim 10~{\rm nm}$) was induced by placing it on h-BN~\cite{butterflyUS,butterflyUK,hunt_arxiv_2013}.

The ever-increasing toolbox of nano-fabrication methods allows today a large flexibility in realizing high quality 2D patterns with nanoscale dimensions in semiconductor quantum structures hosting ultra-high mobility electrons.  An external potential landscape with honeycomb geometry that acts as a lattice of potential wells (such as quantum dots) to trap electrons and/or holes can be obtained by a combination of e-beam nano lithography, reactive ion etching and deposition of metallic gates. The spatial resolution of these techniques can reach values of a few tens of nanometer or even below. Further improvements in spatial resolution can be obtained by bottom-up nanofabrication methods, {\it e.g.} by designing semiconductor lattices by nanocrystal self assembly~\cite{evers_nanolett_2013}. The possibility of independently controlling the electron density and inter-site distances allows one to tune the interplay between on-site ($U$) and nearest-neighbor ($V$) repulsive interactions and single-particle hopping ($t$), opening the way to the observation of collective phenomena and quantum phase transitions in such AG solid-state systems.

Recent experimental results obtained in honeycomb patterns defined on 2DEGs in GaAs quantum heterostructures offer exciting evidence that AG in semiconductors can be realized in the laboratory~\cite{gibertini_prb_2009,desimoni_apl_2010,singha_science_2011,nadvornik_njp_2012}. Available experimental results in the regime $U/t \gg 1$ reveal~\cite{singha_science_2011} unique low-lying collective excitations, such as ``anomalous spin waves", in spectra of inelastic light scattering (see also below and Fig.~\ref{figurefive}a-c). Theoretical studies~\cite{park_nanoletters_2009,gibertini_prb_2009,nadvornik_njp_2012} indicate that Dirac bands can be designed to occur under realistic conditions. For example, elementary tight-binding calculations indicate that Dirac cones extending for $1~{\rm meV}$ can be obtained by tuning the quantum dot spacing to $20~{\rm nm}$, a value reachable by state-of-the-art top-down nanofabrication methods.

\begin{figure*}[t]
\centering
\includegraphics[width=0.8\linewidth]{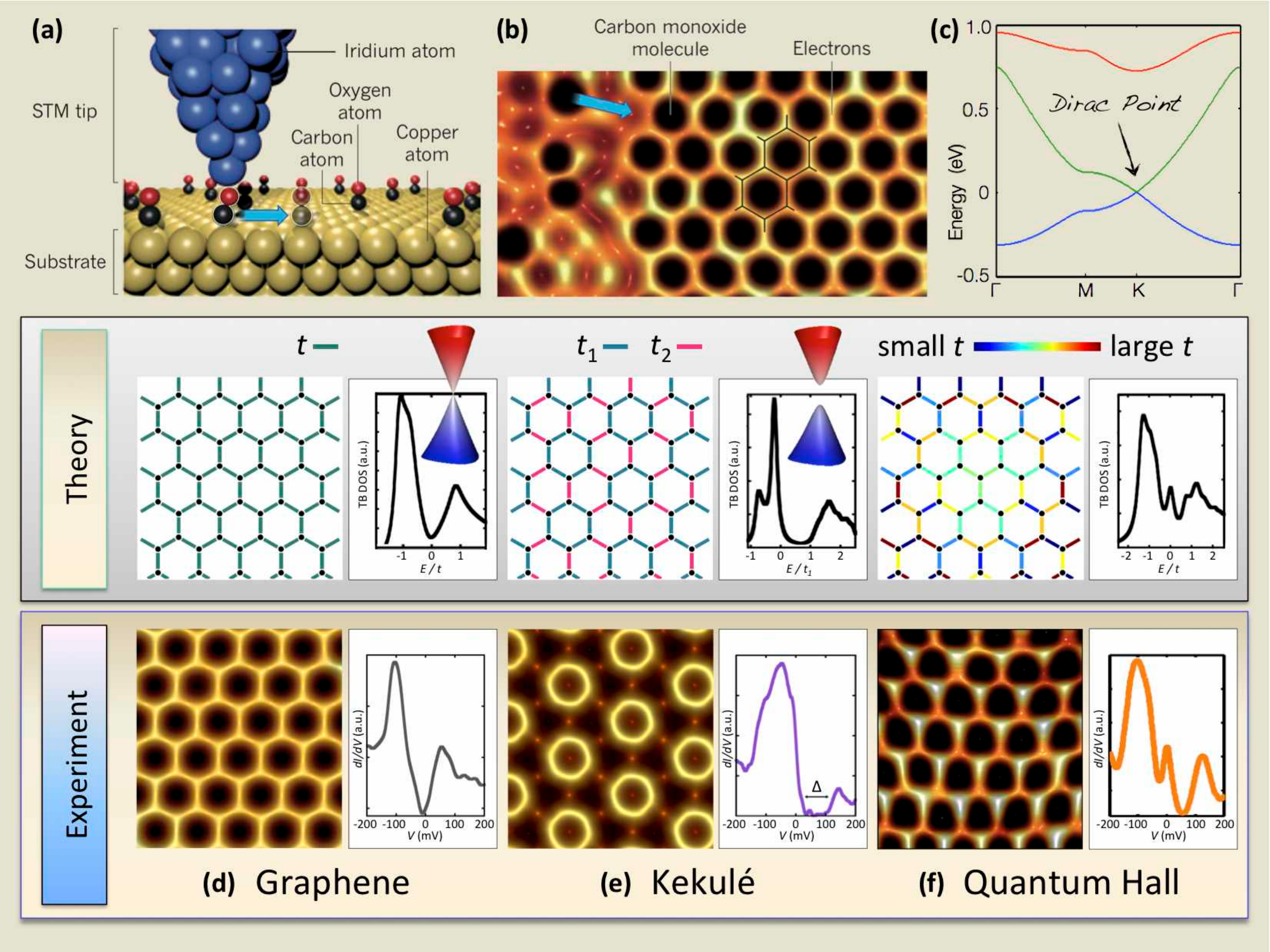}
\caption{{\bf Designer Dirac fermions in molecular graphene}. 
(a) Artificial ``molecular" graphene~\cite{gomes_nature_2012} is fabricated via atom manipulation, and then imaged and locally probed via scanning tunneling microscopy (STM).  Carbon monoxide (CO) molecules are individually positioned (blue arrow) with the STM tip into a triangular lattice on a Copper substrate.  (b) STM topograph of molecular graphene during assembly, showing 2D surface-state electrons repelled from the molecules and guided into a honeycomb lattice (black lines).  (c) Band structure of molecular graphene calculated using the nearly-free electron (NFE) model applied to the periodically-perturbed surface electron gas~\cite{park_nanoletters_2009}.  The relevant parameters have been chosen to match the experimental results in Ref.~\cite{gomes_nature_2012}.  A Dirac band crossing appears at the K point of the superlattice Brillouin zone.  Similar band structures can also be obtained in the tight-binding limit~\cite{park_nanoletters_2009,gibertini_prb_2009} 
and mapped to the NFE model~\cite{gomes_nature_2012}. Progressively more exotic variants of graphene have been fabricated using this method~\cite{gomes_nature_2012}: (d) pristine quasi-neutral graphene exhibits emergent massless Dirac fermions ($d = 19.2~{\rm \AA}$, $t = 90~{\rm meV}$, $t' = 16~{\rm meV}$); (e) graphene with a Kekul\'e distortion ($t_1 = 2 t_2$) dresses the Dirac fermions with a scalar gauge field creating mass ($0.1 \pm 0.02~m_{\rm e}$, $m_{\rm e}$ being the bare electron mass in vacuum); (f) graphene with a triaxial strain distortion embeds a vector gauge field condensing a time-reversal-invariant relativistic quantum Hall phase (shown here for a large pseudo-magnetic field of $60~{\rm Tesla}$). In the theory panel, images are color representations of the strength of the effective Carbon-Carbon bonds (corresponding to tight-binding hopping parameters $t$), and the curves shown are calculated electronic density of states (DOS) from tight-binding (TB) theory. Insets show gapless and gapped Dirac cones matching the experimental data.  In the experiment panel, images are STM topographs acquired after molecular assembly ($100~{\rm \AA}$ field of view, $T = 4.2~{\rm K}$), and the curves are normalized tunnel conductance spectra obtained from the associated nanomaterial.\label{figuretwo}}
\end{figure*}

A completely different route to realize AG with solid-state materials has been recently followed in Ref.~\cite{gomes_nature_2012} (Fig.~\ref{figuretwo}). These authors succeeded in making AG structures with a lattice constant of a few nm by placing CO molecules on top of a Cu substrate with the aid of the tip of a scanning tunneling microscope (STM)~\cite{gomes_nature_2012} (Fig.~\ref{figuretwo}a,b). By probing the density of states of the confined electrons via STM measurements (Fig.~\ref{figuretwo}d) and their evolution as a function of an applied pseudo-magnetic field they proved the formation of Dirac bands with the characteristic Landau levels of Dirac fermions~\cite{NGPNG09,katsnelsonbook}.  While the screening exerted by the bulk states underneath the 2DEG on the Cu(111) surface makes these ``molecular graphene" structures not ideal candidates for exploring many-body effects, the large versatility in the atomic design allows unprecedented local control to embed, map, and tune the symmetries underlying the 2D Dirac equation.  In this system, the authors estimate $U/t \sim 0.5$ using known material parameters~\cite{solovyev_prl_1998,gomes_nature_2012}.  Within the Cu system, there is room to increase the strength of effective interactions by reducing $t$ with larger lattices, or else the atomic manipulation scheme may be extended to other substrates with lower screening effects in a quest to realize other interacting phases (see below).

The band structure of molecular graphene can be understood by assuming that the superlattice potential created by the CO molecules acts as a weak perturbation on the parabolic band that describes the Cu surface state. The superlattice potential is most effective at changing the parabolic dispersion at the edges of the superlattice BZ. Results of such a perturbative calculation \cite{park_nanoletters_2009} are shown in Fig.~\ref{figuretwo}c.  The superlattice potential hybridizes the six unperturbed Cu surface states which lie at the corners of the new BZ. The effective Hamiltonian at each corner of the BZ is given by a $3 \times 3$ matrix, which gives rise to a doublet (blue and green bands in Fig.~\ref{figuretwo}c) and a singlet (red band in Fig.~\ref{figuretwo}c). The doubly degenerate state leads to an effective Dirac equation.  This ``nearly free" electron scheme can be generalized to include the effect of strain~\cite{gomes_nature_2012} and spin-orbit coupling~\cite{ghaemi_prb_2012}.  Similar approaches can be applied to describe the appearance of Dirac bands in AG in semiconductors \cite{park_nanoletters_2009,gibertini_prb_2009}. The general symmetries of the triangular lattice uniquely determine these couplings, which have the same form as graphene~\cite{KM05b,M07}.  In particular, spatially patterning the hopping via STM atom manipulation allows the generation of both gauge (pseudo) electric and magnetic fields in molecular graphene (see Fig.~\ref{figuretwo}e,f).  Global changes in the lattice constant in molecular graphene add a simple scalar potential to the Dirac Hamiltonian equivalent to an electrical field which changes the chemical potential or ``doping"~\cite{gomes_nature_2012}.  Local changes to the lattice constant engineer a strain introducing a vector potential equivalent to a large perpendicular magnetic field~\cite{gomes_nature_2012,GKG10,levy_science_2010} here tunable up to $60~{\rm Tesla}$ (Fig.~\ref{figuretwo}f).  Finally, creating an alternating bond structure in the form of a Kekul\'e distortion was shown (Fig.~\ref{figuretwo}e) to attach mass to the formerly massless Dirac fermions, akin to the Higgs field~\cite{jackiw_aip_2007,hou_prl_2007,roy_prb_2010}.

For future experiments, the availability of semiconductors (InAs, InSb, etc.) and metals (Ag, Au, etc.) with large spin-orbit coupling 
creates concrete and exciting possibilities to explore topological phases of artificial matter with these approaches~\cite{gibertini_prb_2009,zhang_prb_2011, ghaemi_prb_2012,sushkov_arxiv_2012}.

\subsection{Confining photons}

Photonic crystals, an optical
analogue of ordinary crystals, offer an additional route to design energy dispersion 
relations with characteristic Dirac points~\cite{Parimi04}. 
In such crystals, the unusual transmission properties
near a Dirac point~\cite{haldane_prl_2008,sepkhanov07,sepkhanov08} were predicted and observed experimentally.

The state of the art of photonic crystals operating in the microwave frequency range is well described in Ref.~\cite{Bittner10}. 
In this work the crystal is 2D and composed
of rows of metallic cylinders, which are arranged to form a
triangular lattice. Electromagnetic waves propagating in such a
periodic structure, composed of metallic cylinders with radius 
$R = 0.25~a$, where $a$ is the lattice constant, exhibit a dispersion relation with
several Dirac points. In the
vicinity of a Dirac point, the measured reflection spectra
resemble the STM spectra of graphene flakes~\cite{STMgraphene1,STMgraphene2}. 
In a subsequent work~\cite{Bittner12} extremal transmission through a microwave
photonic crystal and the observation of edge states close to Dirac
points were also demonstrated. The authors of Ref.~\cite{Bittner12} have shown that 
the transmission through this crystal displays a pseudo-diffusive~\cite{sepkhanov07} $1/L$ dependence on the thickness $L$ 
of the crystal. In addition, they measured the eigenmode intensity
distributions in a rectangular microwave billiard that contains a triangular photonic crystal. 
Close to the Dirac point there appear states at the straight edge of the photonic crystal that
represent the artificial counterpart of the states at a zigzag edge of natural graphene. 
Optical analogues of graphene operating in the microwave frequency range have been recently used 
to simulate {\it anisotropic} honeycomb lattices and to observe topological phase transitions of Dirac points~\cite{bellec_prl_2013}.

\begin{figure*}[t]
\centering
\includegraphics[width=0.60\linewidth]{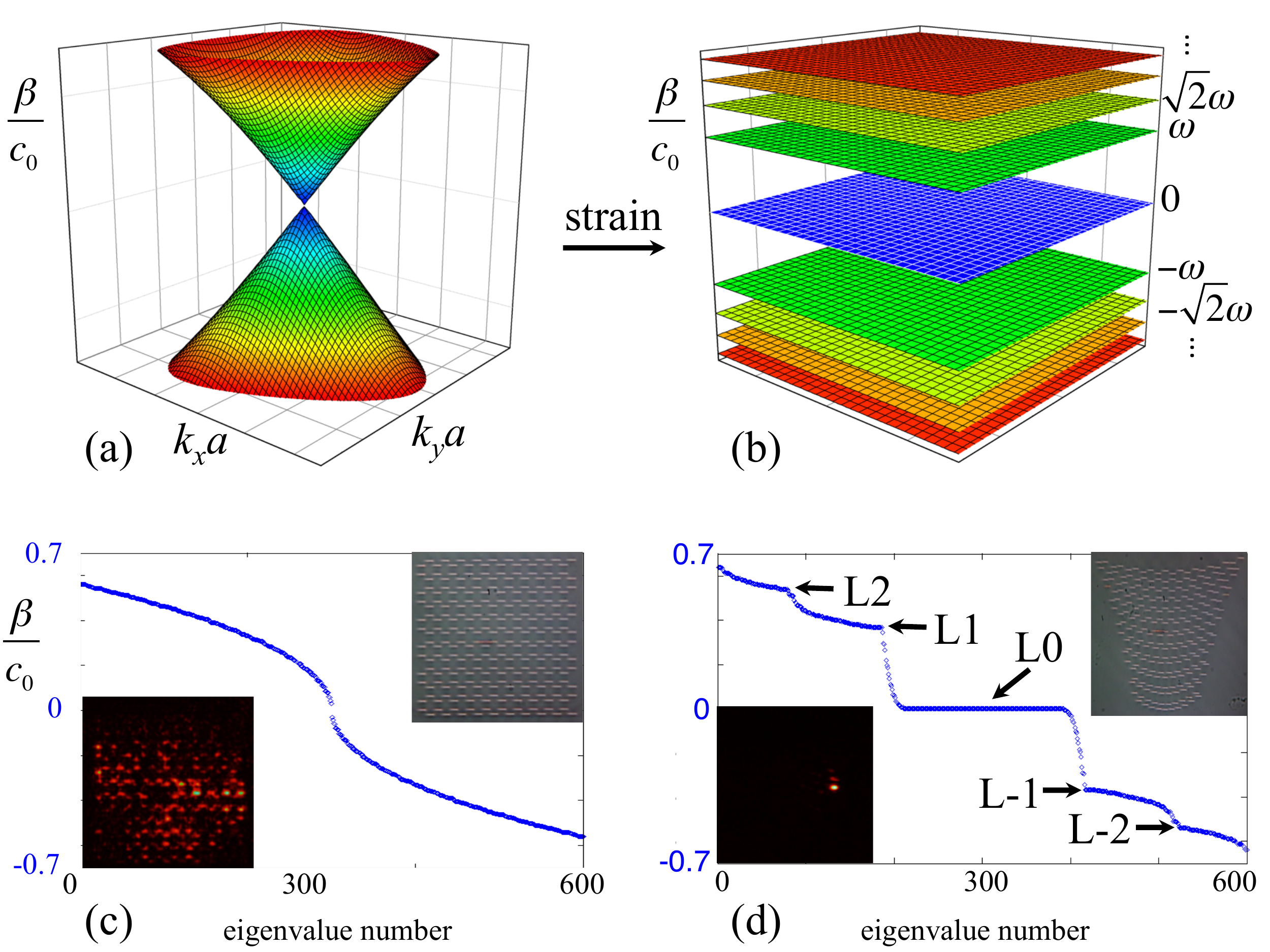}
\caption{{\bf Landau levels of photons in strained photonic graphene}. (a) Dirac cone in
spatial spectrum of unstrained photonic graphene, as obtained by solving Eq.~(\ref{schroedingerlike}) with the Ansatz $\Psi(x,y,z)  = \phi(x,y)\exp(\beta z)$. As in Fig.~\ref{figureone}, $c_0$ plays the role of the hopping parameter $t$. 
(b) Spectrum of inhomogeneously strained photonic graphene, according to a strain tensor
that gives a constant pseudo-magnetic field.  The planes are highly
degenerate photonic Landau levels~\cite{Rechtsman12a};  (c) Eigenvalues in the unstrained
lattice are listed in descending order.  Insets show a microscope image of
the photonic lattice (top right) and strong diffraction of light through
the lattice (bottom left);  (d) Eigenvalues in the strained lattice listed
in descending order, showing clear evidence of Landau levels (labeled by ``${\rm L}n$" with $n=-2,-1,0,+1, +2, {\rm etc}$).  Microscope image of the strained lattice is shown (inset top right),
as well as resulting experimental output showing strong optical confinement
in Landau level band gaps (bottom left). Courtesy of M. Segev and M. Rechtsman. \label{figurethree}}
\end{figure*}

Since 2007 intensive studies of honeycomb lattices constructed by the method of optical
induction have been performed. 
The honeycomb structure in Ref.~\cite {Peleg07} was induced by the
intensity pattern $I(x,y)$ of three interfering plane waves,
which is translated into a change in the refractive index  $\Delta n$
through the nonlinearity in a photorefractive crystal. 
Such a lattice exhibits several Dirac points, which have been termed in Ref.~\cite {Peleg07} {\it diabolical points} after M.V. Berry and M. Wilkinson.
The paraxial evolution of the complex amplitude $\Psi$ of a
probe beam propagating in the lattice is governed by a
normalized Schr\"{o}dinger-type equation of the form:
\begin{equation}\label{schroedingerlike}
i\frac{\partial \Psi}{\partial z}+ \nabla^2_{\perp}- \frac{V_0\Psi}{1 + I(x,y)+ |\Phi|^2}=0~,
\end{equation}
where $V_0$ controls the relative values of potential depth and
nonlinearity strength. The wave dynamics in such honeycomb photonic lattices 
has been extensively studied~\cite{Peleg07} offering evidence 
of the unique phenomenon of {\it conical diffraction} around
the singular diabolical (Dirac) points connecting the first and second bands. 
These findings have represented the first experimental observation of the phenomenon of conical diffraction,
predicted by W.R. Hamilton in the 19th century, arising here
solely from a periodic potential. In addition, ``honeycomb gap solitons", residing in the gap between the second and the third band, were observed, reflecting yet another special property of honeycomb photonic lattices.

In recent years several theoretical papers were published concerning various aspects of the physics of 
honeycomb photonic lattices. These include studies of i) $PT$-symmetry 
({\it i.e.} symmetry under the combined operations of parity and time reversal), ii) nonlinear wave dynamics, 
iii) the persistence of the Klein effect, and iv) the breakdown of conical diffraction due to nonlinear interactions~\cite{bahattreidel_pra_2010,szameit_pra_2011,bahattreidel_pra_2011}. 
Perhaps the culmination of these studies has been 
presented in Refs.~\cite{Rechtsman12,Rechtsman12a,Rechtsman12b}. 
Ref.~\cite{Rechtsman12} describes combined theoretical and experimental work
on the creation and destruction of topological edge states in ``optical graphene", where, after the application of uniaxial strain (compression), two Dirac points merge resulting in the formation of a band
gap. Effectively, edge states are created (destroyed) on the zig-zag (``bearded") edge
of the structure. Moreover, the authors of Ref.~\cite{Rechtsman12} have claimed the observation of a novel type of ``bearded"
edge state, which cannot be explained by the standard tight-binding theory, while they can be classified as Tamm states lacking any surface effect. This is an example that highlights how AG structures might provide insights on physics beyond that displayed by natural graphene. A second complementary work~\cite{Rechtsman12a} demonstrates the creation of synthetic magnetic fields and ``photonic Landau levels" separated by bandgaps in the spatial spectrum of the structured dielectric lattice, as illustrated in Fig.~\ref{figurethree}. This is the photonic analog of the relativistic electron Landau levels observed in strained molecular graphene~\cite{gomes_nature_2012} (Fig.~\ref{figuretwo}f). 
Finally, a photonic Floquet TI based on an AG structure of helical waveguides, evanescently coupled to one another, has been proposed in Ref.~\cite{Rechtsman12b}. 2D photonic TIs based on optical spin-orbit coupling---achieved through the employment of a metamaterial composed of split-ring resonators---have been proposed in Ref.~\cite{khanikaev_arxiv_2012}.

\subsection{Confining atoms and ions}

Ultracold atoms and ions represent particularly excellent arenas for the field of quantum simulation. 
Efficient quantum simulators of various systems (like Bose- or Fermi-Hubbard models, lattice spin models, etc.) have either been
realized or are within reach in the time span of a few years~\cite{Lewenstein12}. 

In recent years, a great deal of attention was devoted to simulations involving  honeycomb optical lattices.
The main target of these efforts was not to reproduce the physics of natural graphene, but rather to seek out novel phenomena that are beyond reach in Carbon-based 2D honeycomb crystals. Optical lattices are, for example, {\it flexible} in the sense that their geometry can be changed {\it in situ}. Cold atoms in optical lattices, moreover, can be forced into regimes of parameters or subject to external fields, which are difficult to achieve in or unaccessible to natural graphene. Examples include the regime of ultrastrong spin-orbit coupling~\cite{Goldman10} and non-Abelian gauge fields~\cite{Bermudez10,Hauke12} akin to those that appear in the Langrangian of quantum chromodynamics. Note that with the advent of molecular graphene and Abelian gauge fields~\cite{gomes_nature_2012}, proposals now also exist for realizing non-Abelian gauge fields in solid-state incarnations~\cite{dejuan_prb_2013}. Last, but not least, ultracold gases offer novel means to control the nature, strength, and range of inter-particle interactions~\cite{Lewenstein12}.

\begin{figure*}[h!]
\includegraphics[width=0.60\linewidth]{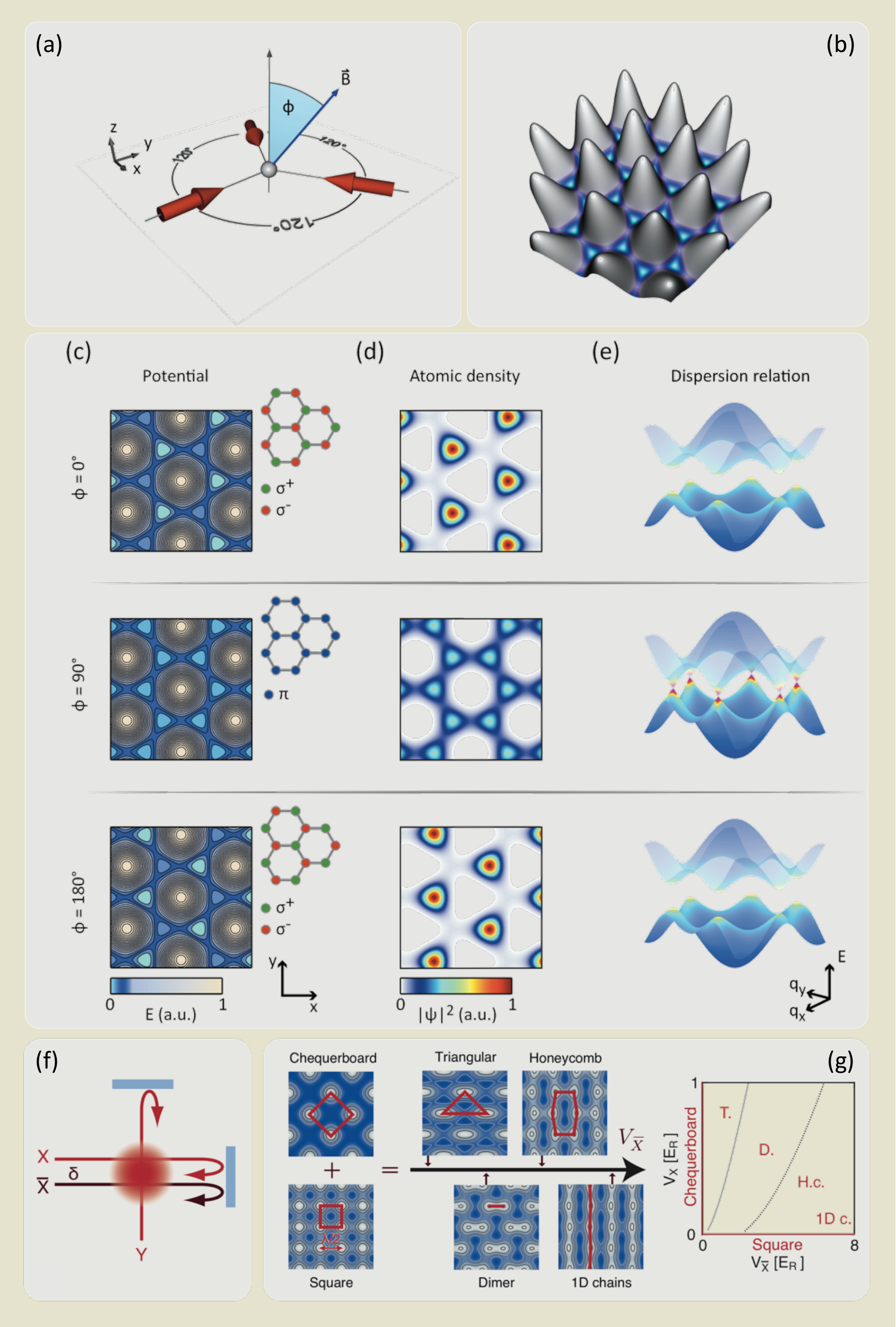}
\caption{{\bf Flexible optical lattices for cold atom gases}. 
(a) Experimental setup for a spin-dependent hexagonal lattice~\cite{SoltanPanahi11a}: three running laser beams intersect at an angle of 
$120^{\circ}$ with respect to each other. The magnetic field ${\bm B}$ forms an angle $\phi$ with the ${\hat {\bm z}}$ azis. 
(b) Resulting scalar potential in the dipole approximation. 
(c) The total potential for $^{87}{\rm Rb}$ atoms in the Zeeman state $|F= 1, m_{\rm F} = +1\rangle$ is plotted for $\phi = 0^{\circ}, 90^{\circ}$, and $\phi = 180^{\circ}$. Insets show the corresponding light polarization on the two sublattices. 
(d) Atomic density distributions in the lowest Bloch band of the potentials plotted in c). 
(e) Dispersion relation of the two lowest-energy bands. The colormap illustrates the curvature of the bands. Note that the results for  $\phi=180^{\circ}$ can be obtained from  those of  $\phi=0^{\circ}$ by exchanging minima and maxima of the total potential and atomic density, as well as red and green bullets. 
(f) An optical lattice of tunable geometry has been created in Ref.~\cite{Tarruell12} by using three retro-reflected laser beams. Beams $X$ and $Y$ interfere and produce a chequerboard pattern, while beam ${\bar X}$ creates an independent standing wave. Their relative position is controlled by the detuning $\delta$. 
(g) A large range of lattices can be realized depending on the intensities of the lattice
beams. They result from the overlap of chequerboard and square-lattice patterns. White (blue) regions denote lower (higher) values of the potential energy. The diagram on the right shows the
accessible lattice geometries as a function of the lattice depths $V_{\bar X}$ and
$V_X$ associated with beams ${\bar X}$ and $X$, respectively. 
The transition between triangular (``T") and dimer (``D") lattices is
indicated by a dotted line. Dirac points appear in the 
honeycomb regime (``H.c."), {\it i.e.} to the right of the dashed line. The limit $V_{\bar X}\gg V_X,V_Y$
corresponds to weakly coupled, one-dimensional chains (``1D c."). 
Panels (a)-(e) courtesy of M. Weinberg and K. Sengstock. Panels (f)-(g) courtesy of L. Tarruell, T.  Uehlinger, and T. Esslinger.\label{figurefour}}
\end{figure*}

The authors of Ref.~\cite{SoltanPanahi11a} have pioneered attempts to create flexible honeycomb lattices. In this work the first realization of an ultracold ($^{87}{\rm Rb}$) Bose gas in a spin-dependent optical lattice with hexagonal symmetry was reported. The basic structure of this lattice is illustrated in Figs.~\ref{figurefour}a-e. Three linearly-polarized laser beams intersect at an angle of $120^{\circ}$, yielding local potential minima in an hexagonal geometry (Fig.~\ref{figurefour}a,b) and a local polarization that alternates between $\sigma^+$ and $\sigma^-$ when one goes from one sublattice to the other. 
As atoms in a light field experience a polarization-dependent ac Stark shift, the potential at  $\sigma^+$ and $\sigma^-$ sites is different for different atomic Zeeman substates labeled by $m_{\rm F}$. The angle between the ${\hat {\bm z}}$ axis and a homogenous static magnetic field ${\bm B}$, which defines the quantization axis, is denoted by $\phi$ (Fig.~\ref{figurefour}a). According to the local polarization, the two interpenetrating triangular lattices forming the hexagonal lattice are denoted as ``$\sigma^+$" and ``$\sigma^-$" lattices. The hexagonal lattice can therefore also be regarded as a triangular lattice with a basis where the atoms occupy the $\sigma^+$ and $\sigma^-$ sites indicated in Fig.~\ref{figurefour}c  for  $\phi=0^{\circ}$ by green and red bullets. In contrast, for  $\phi=90^{\circ}$ the lattice is perfectly hexagonal (and the asymmetry between the two sublattices vanishes).

For $\phi=0^{\circ}, 180^{\circ}$ the effective potential felt by an atom can be written as:
\begin{equation}\label{eq:potentialHamburg}
  V({\bm x}) = V_{\rm hex}({\bm x}) + m_{\rm F} g_{\rm F} \mu_{\rm B} B_{\rm eff}({\bm x})~, \label{potential}
\end{equation}
where the polarization of the light field ${\cal P}({\bm x})$---where ${\cal P}({\bm x}) = \pm 1$ for pure $\sigma^\pm$ polarizations---is mapped onto a pseudo-magnetic field $B_{\rm eff}({\bm x})\propto -V_{\rm hex}({\bm x}){\cal P}({\bm x})/\mu_{\rm B}$, where $g_{\rm F}$ is the Land\'e 
$g$-factor and $\mu_{\rm B}$ the Bohr magneton. The potential consists of a spin-independent part $V_{\rm hex}({\bm x})$ of hexagonal symmetry and a state-dependent superlattice emerging from the local pseudo-magnetic field $B_{\rm eff}({\bm x})$. The single-particle energy-momentum dispersion relation of an atom evolves as $\phi$ changes from a gapped to a gapless one exhibiting Dirac points (Fig.~\ref{figurefour}e).
Note that the maxima of the potential at the center of each honeycomb cell prevent direct diagonal 
tunneling through these maxima. The interplay between single-particle band structure effects and interactions between atoms leads to interesting many-body physics~\cite{SoltanPanahi11a}, 
which will be discussed in the next Section. 

\begin{figure*}[h!]
\centering
\includegraphics[width=0.6\linewidth]{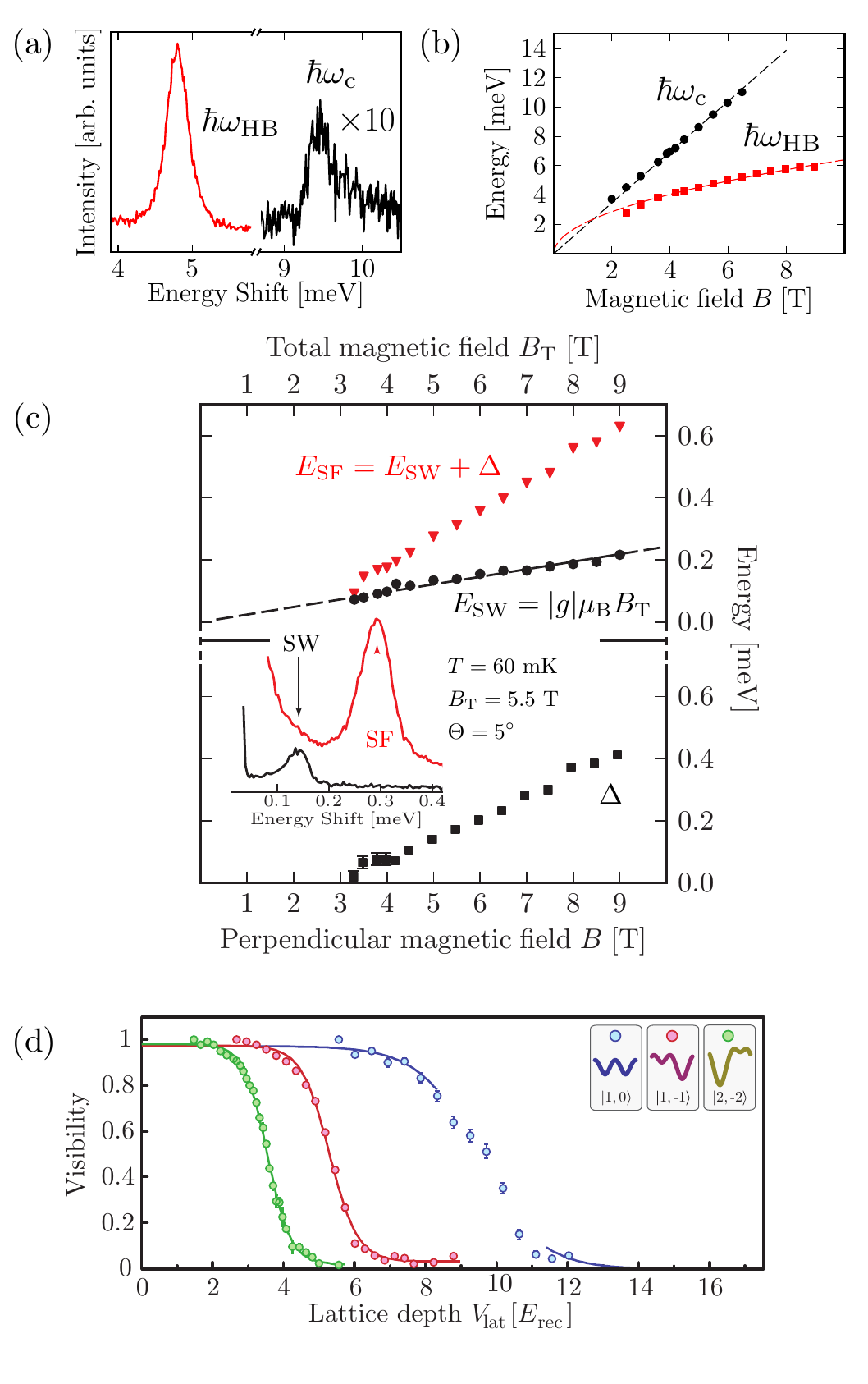}
\caption{{\bf Many-body effects in artificial graphene}. 
(a) Resonant inelastic light scattering spectra (at $B = 5.48~{\rm Tesla}$ and $T = 1.7~{\rm K}$) of a high-mobility 2D electron gas in the presence of a honeycomb lateral superlattice~\cite{singha_science_2011} showing a weak peak in the excitation spectrum due to the ordinary cyclotron mode and a much stronger peak due to the Hubbard mode. (b) Evolution of the energies of the cyclotron mode (black filled circles) and of the Hubbard mode (at frequencies $\omega_{\rm HB}$) (red filled squares) at $T = 1.7~{\rm K}$. The  black dashed line is a linear fit to the data, $\hbar \omega_{\rm c} = \hbar eB/(m^{*}c)$, with $m^{*} = 0.067~m_{\rm e}$. The red dashed line is a fit of the type $\hbar\omega_{\rm HB} = \alpha \sqrt{B[{\rm Tesla}]}$ with $\alpha \sim 2~{\rm meV}$. (c) Energies of two spin collective modes: the ordinary spin-wave mode (black filled circles) and an anomalous ``spin-flip" mode (red filled triangles). 
The black dashed line is a linear fit to the data, $E_{\rm SW} = |g| \mu_{\rm B} B_{\rm T}$, with $|g| = 0.42$. 
Representative excitation spectra of both spin modes at two different laser energies are reported in the inset 
($B_{\rm T}$ is the total magnetic field and $\theta$ the tilt angle of the sample with respect to the magnetic field).
The black filled squares label the splitting $\Delta$ between the two spin modes. (d) Superfluid-to-Mott-insulator transition in a spin-dependent honeycomb optical lattice~\cite{SoltanPanahi11a}. The visibility of the interference fringes in a time-of-flight experiment is plotted as a function of the lattice depth (in units of the recoil energy). At the superfluid-to-Mott-insulator transition the visibility drops down considerably. Data labeled by blue symbols refer to the superfluid-to-Mott-insulator transition in a  spin-{\it independent} hexagonal lattice, {\it i.e.} $m_{\rm F} =0$ in Eq.~(\ref{eq:potentialHamburg}). The two other curves refer to finite values of $m_{\rm F}$, as indicated in the inset. Panels (a)-(c) courtesy of M. Gibertini (experimental data from Ref.~\cite{singha_science_2011}). Panel (d) courtesy of C.  \"{O}lschl\"{a}ger and P. Windpassinger (experimental data from Ref.~\cite{SoltanPanahi11a}). \label{figurefive}}
\end{figure*}

The authors of Ref.~\cite{Tarruell12} have carried out pioneering work on cold ($^{40}{\rm K}$) Fermi gases in honeycomb optical lattices. They have created, moved, and merged Dirac points in a tunable honeycomb lattice. More recently, the same authors have studied double
transfer through Dirac points in a honeycomb optical lattice~\cite{Uehlinger12}.  More precisely, these authors measured the quasi-momentum distribution of the atoms after sequentially passing through two Dirac points. They observed a double-peak feature in the fraction of atoms transferred to the second band, both as a function of the band gap at the Dirac points and the quasi-momentum of the trajectory. They could not observe coherent St\"uckelberg oscillations due to the variation of the potential gradient over the atomic cloud size. The flexibility of the lattice used in these experiments is illustrated in  Fig.~\ref{figurefour}f,g. The laser configuration is based on three retro-reflected beams,  which by control of the detuning $\delta$, create a checkerboard, a triangular, a honeycomb lattice, or a lattice of weakly coupled 1D chains. These fermionic optical lattice techniques~\cite{Tarruell12} were recently contrasted~\cite{simon_nature_2012} with complementary methods for assembling electron lattices in molecular graphene~\cite{gomes_nature_2012}. 

More work on cold atoms confined in flexible optical lattices will be reviewed in the next Section.

Trapped ions provide perhaps the best quantum controlled systems
available nowadays~\cite{blatt_naturephys_2012}. However, their application for quantum
simulations of AG has been so far restricted only to small systems. Experimental 
studies on trapped ion systems focused so far on the simulation of the Dirac
equation encoded in motional and internal degrees of freedom of a
trapped ion in 3+1 and 1+1 space-time dimensions, 
and its consequences, such as {\it Zitterbewegung}~\cite{Gerritsma10} and the Klein paradox~\cite{Gerritsma11}. 
There is, however, a large ongoing effort toward scalable ion systems, employing for instance arrays of microtraps or 
self-assembled lattices, where several hundreds of ions in a triangular lattice can be stored~\cite{Britton12}.

\section{Strongly correlated phenomena and topological phases in honeycomb lattices}

The most interesting phases of matter, including ferromagnetism and superconductivity, 
emerge in systems with a macroscopically large number of degrees of freedom in the presence of interactions.
Here, we discuss how some prominent broken-symmetry states can be explored in artificial honeycomb lattices. 

The great advantage of these artificial crystals is that interactions among electrons, atoms, ions, and photons 
can be engineered and largely tuned. In artificial semiconductor lattices, for example,  
one can quench the kinetic energy by applying a magnetic field perpendicular to the 2D electron system, thereby emphasizing the role of long-range Coulomb interactions ~\cite{park_nanoletters_2009,gibertini_prb_2009,desimoni_apl_2010,singha_science_2011}. In an ultracold atomic gas system, the strength of interactions between atoms can be tuned at will by employing Feshbach resonances~\cite{duine_physrep_2004,Lewenstein12}. Finally, significant photon-photon interactions can be achieved by employing nonlinear optical media, thereby offering the possibility to realize strongly interacting fluids of light~\cite{carusotto_arxiv_2012}.

\subsection{Hubbard correlations and split bands}

The Hubbard model~\cite{hubbardI,hubbardIII} is the cornerstone of the physics of strongly correlated systems. 
The Hubbard Hamiltonian encodes a daunting competition between two energy scales~\cite{kotliar_pt_2004,kotliar_rmp_2006,vollhardt_aip_2010}: the kinetic energy $t$, which measures the overlap between wave functions on neighboring lattice sites, and the interaction energy $U>0$, which measures the strength of on-site repulsions. The single-band Fermi-Hubbard Hamiltonian reads:
\begin{equation}\label{Hubbard}
{\hat {\cal H}} = - t \sum_{\langle i,j\rangle, \sigma} {\hat c}^\dagger_{i, \sigma} {\hat c}_{j, \sigma} 
+ U \sum_i {\hat n}_{i, \uparrow} {\hat n}_{i, \downarrow}~.
\end{equation}
Here ${\hat c}^\dagger_{i, \sigma}$ (${\hat c}_{i, \sigma}$) creates (destroys) a fermion with spin $\sigma$ at site $i$ of the honeycomb lattice and ${\hat n}_{i, \sigma} = {\hat c}^\dagger_{i, \sigma} {\hat c}_{i, \sigma}$ is the spin-resolved number operator. In the first term, the sum is over all pairs $\langle i,j \rangle$ of nearest-neighbor sites. A straightforward generalization of the Hamiltonian (\ref{Hubbard}) that applies to bosons exists and it is usually termed the ``Bose-Hubbard model"~\cite{fisher_prb_1989,fazio01}.

When interactions are negligible ($U/ t \ll 1$), the ground state of the honeycomb-lattice Fermi-Hubbard Hamiltonian is semimetallic and characterized by linearly-dispersive massless-Dirac-fermion conduction and valence bands touching at two inequivalent points in the BZ. In the non-perturbative regime, $U/t \gtrsim 1$, the ground-state phase diagram of this model has been extensively studied by means of quantum Monte Carlo (QMC) techniques~\cite{kotov_rmp_2012}. Early on it was shown that at half filling and for $U/t \gtrsim 5$ a semimetal-Mott insulator transition occurs~\cite{sorella_epl_1992}. Recently, a QMC calculation has demonstrated~\cite{meng_nature_2010} the existence of a gapped anti-ferromagnetic phase for $U/t > 4.3$ and presented evidence for a gapped spin liquid phase in the range of couplings $3.5 < U/t < 4.3$. The findings of Ref.~\cite{meng_nature_2010} have been recently addressed in Ref.~\cite{sorella_scirep_2012}, by QMC simulations of the same model in larger clusters (containing up to $2592$ sites), finding very weak evidence of a spin liquid phase. 

An ``extended" Hubbard model involving an additional interaction term describing nearest neighbor repulsions of strength $V$ has been studied analytically by employing renormalization-group (RG) techniques in the limit of a large number $N_{\rm f}$ of fermion flavors~\cite{herbut_prl_2006,herbut_prb_2009} (we remind the reader that $N_{\rm f} =4$ for electrons in graphene). It has been found~\cite{herbut_prl_2006,herbut_prb_2009} that sufficiently large values of $V/t$ stabilize charge-density-wave phases over semimetallic or Mott insulating phases. Numerical RG calculations~\cite{raghu_prl_2008} have qualitatively confirmed these results but also discovered that second-neighbor repulsions favor topological states with spontaneously broken time-reversal symmetry ({\it i.e.} quantum spin Hall phases) over charge- or spin-density waves. 

Finally, it has been recently shown~\cite{nandkishore_naturephys_2012} that, when the Fermi energy is moved away from the Dirac point and the system is doped to the vicinity of a van Hove singularity in the density of states (such as the one that occurs at the $M$ point of the honeycomb-lattice band structure~\cite{NGPNG09,kotov_rmp_2012}), repulsive short-range interactions favor $d + id$ chiral superconductivity over many other competing orderings. An alternative route to non-conventional superconductivity in AG and related compounds relies on the suppression of long-range forces, while keeping a large value ($U/t \sim 1$) of the Hubbard on-site repulsion~\cite{GU12,RCG13}.

All these predictions call for future experimental efforts. To this end we note that  strong correlations leave deep scars on the excitation spectrum of a many-body system, which can be probed by a variety of spectroscopic tools. The simplest example is represented by a gapped collective mode between ``Hubbard split bands". In the strongly correlated or ``atomic" $U/t\gg 1$ limit the Hubbard model (\ref{Hubbard}) displays two bands~\cite{hubbardIII, hansen_prb_2011}, which are split by the energy cost $U$ of having two fermions with antiparallel spin on the same site. In the atomic limit it is therefore natural to expect a gapped collective mode in which particles belonging to the lower Hubbard band are cooperatively promoted to the upper Hubbard band. Such a mode has been observed in a Bose-Einstein condensate in a deep 3D optical lattice~\cite{greiner_nature_2002} and, more recently, in an artificial honeycomb lattice realized by nanopatterning the surface of a GaAs semiconductor~\cite{singha_science_2011} (see Fig.~\ref{figurefive}). This AG system displays also an anomalous ``spin-flip" mode (Fig.~\ref{figurefive}c) that could arise from the removal of the sublattice-pseudospin degeneracy due to inter-site Coulomb repulsions~\cite{alicea_prb_2006}. Explorations of Hubbard excitations and anomalous spin-flip modes can be extended to molecular and photonic AG once interactions in these systems are turned on.

As we have already stated above, a plethora of many-body effects and lattice models can be very effectively simulated by employing cold atom gases~\cite{Lewenstein12}. For example, in Ref.~\cite{SoltanPanahi11a} combined effects of the lattice and inter-atomic interactions led to a forced antiferromagnetic N\'eel order, in which two spin-components localize at different lattice sites. Coexistence of Mott-insulator- and superfluid-type orders leads to the formation of a forced supersolid. Next-nearest-neighbor tunneling seems to play a role in the physics described in Ref.~\cite{SoltanPanahi11a}, a fact that paves the way for the realization of the famous Haldane model~\cite{H88} in this system. Later on, the authors of Ref.~\cite{SoltanPanahi11b} reported the observation of a quantum phase transition to a multi-orbital superfluid phase in an optical lattice. In this unconventional superfluid, the local phase angle of the complex order parameter is continuously twisted between neighboring lattice sites. The nature of this twisted superfluid quantum phase is an interaction-induced admixture of the $p$-orbital contributions favored by the graphene-like band structure of the hexagonal optical lattice used in the experiment. 

In Ref.~\cite{Struck11} various forms of frustrated classical ferromagnetism have been studied by transforming the lattice geometry 
from square to triangular to an array of linear chains and back.  Finally, in a recent experimental work~\cite{Greif12} 
novel instances of quantum magnetism of ultracold fermions have been reported. In this work short-range magnetic order has been achieved by loading two-component Fermi gases in either a dimerized or anisotropic simple cubic optical lattice. Flexible kagome lattices have been recently studied in Ref.~\cite{Gyu-Boyong12}.

\subsection{Long-range interactions}

Electrons in nanopatterned 2DEGs~\cite{park_nanoletters_2009,gibertini_prb_2009,desimoni_apl_2010,singha_science_2011,nadvornik_njp_2012,goswami_prb_2012} 
offer a natural system to study correlation effects in the presence of long-range Coulomb interactions. The applicability of the Hubbard model (\cite{hubbardI,hubbardIII}) to describe these systems when the Fermi energy is at the Dirac point~\cite{NGPNG09,kotov_rmp_2012} (the density of states vanishes) is questionable because the $1/r$ long-range tail of the Coulomb interaction is not screened.  

In recent years, several efforts have been made~\cite{hands_prb_2008,drut_prl_2009,drut_prb_2009,drut_prb_2009_bis} to describe electrons moving in a honeycomb lattice and interacting through the non-relativistic Coulomb force. In this case the strength of interactions is measured by the dimensionless parameter~\cite{NGPNG09,kotov_rmp_2012} $\alpha_{\rm ee} \equiv e^2/(\epsilon \hbar v_{\rm F})$, where $e$ is the absolute value of the electron's charge, $\epsilon$ is a suitably-defined dielectric constant, and $v_{\rm F}$ is the Fermi velocity. This parameter, which formally resembles the QED fine-structure constant $\alpha = e^2/(\hbar c)$ (where $c$ is the speed of light), plays the role of the 
$U/t$ coupling constant in the Hubbard model~\cite{hubbardI,hubbardIII}.

A excitonic insulating phase has been predicted to occur spontaneously~\cite{drut_prl_2009,drut_prb_2009,drut_prb_2009_bis} at a critical value of the Coulomb coupling constant $\alpha_{\rm ee} \simeq 1.1$ for $N_{\rm f} =4$ fermion flavors. Note that electrons in natural suspended graphene are characterized by $\alpha_{\rm ee} \sim 2.2$ (since, in this case, $\epsilon \sim 1$ and $v_{\rm F} \sim 10^6~{\rm m}/{\rm s}$). Weak-field magneto-transport experiments have been carried out~\cite{elias_naturephys_2011,mayorov_nanolett_2012} in high quality suspended samples (with mobilities of the order of $\sim 10^6~{\rm cm}^2~{\rm V}^{-1}{\rm s}^{-1}$), with carrier density fluctuations and temperatures as low as $10^8~{\rm cm}^{-2}$ and $1~{\rm K}$, respectively. No experimental evidence of a gapped phase has been reported so far in nearly-neutral natural graphene~\cite{elias_naturephys_2011,mayorov_nanolett_2012}. The observed behavior of graphene is more consistent with the existence of a strong renormalization of the Fermi velocity~\cite{elias_naturephys_2011} and an RG flow towards a weakly coupled phase (see, however, Ref.~\cite{mueller_prl_2009}). When one dopes the system away from the neutrality point, screening kicks in and the electron fluid in single-layer graphene behaves as a Fermi liquid, albeit with a number of intriguing twists~\cite{polini_ssc_2007,barlas_prl_2007,kotov_rmp_2012}.

Electrons in artificial honeycomb lattices~\cite{park_nanoletters_2009,gibertini_prb_2009,desimoni_apl_2010,singha_science_2011} therefore offer the unique opportunity to study strong correlation effects, which are beyond reach in natural graphene. The role of long-range electron-electron interactions in artificial graphene has been studied by means of density-functional theory in Ref.~\cite{rasanen_prl_2012}. Electron-electron interactions have been demonstrated~\cite{rasanen_prl_2012} to shift the threshold for the emergence of isolated Dirac points to larger well depths than found without interactions~\cite{park_nanoletters_2009,gibertini_prb_2009}. This effect is particularly pronounced when the number of electrons per well is increased.

Long-range interactions can nowadays also be studied in the realm of atomic physics. Recent experimental advances in cooling atoms with permanent dipole moments and polar molecules have indeed made it possible to study quantum gases with {\it dipolar} interactions decaying like $1/r^3$ at large distances. These systems have attracted a large fraction of theoretical and experimental interest: detailed information can be found in recent review articles~\cite{lahaye09,baranov12}.

\section{Summary and perspectives}

In summary, in this Article we have reviewed recent progress in the creation of artificial graphene focusing on nanopatterning of ultra-high-mobility two-dimensional electron gases in semiconductors, molecule-by-molecule assembly via scanning probe methods, confining photons in dielectric crystals, and optical trapping of ultracold atoms in crystals of light. Lattices of superconducting circuits~\cite{houck_naturephys_2012} may also offer further opportunities to study strongly correlated phases of light in honeycomb structures. Fully tunable {\it plasmonic} analogues of graphene can be realized in two-dimensional honeycomb lattices of metallic nanoparticles~\cite{weick_prl_2013}.

The occurrence of fragile interaction-induced broken-symmetry states in artificial graphene might be preempted by external unwanted random potentials, which induce inhomogeneities in the particle-number distribution. Controlled sources of disorder can, however, be exploited to investigate the interplay between disorder and interactions~\cite{deissler_naturephys_2010} and the occurrence of ``Bose glass"~\cite{fisher_prb_1989} phases.

Finally, we foresee useful applications of artificial graphene for the realization of new topological phases of matter and the engineering of Abelian and non-Abelian gauge fields.

\section{Acknowledgements}
It is a great pleasure to thank Rosario Fazio, Misha Katsnelson, Aron Pinczuk, and Giovanni Vignale for very useful and stimulating discussions. We acknowledge financial support by the Spanish Ministry of Economy (MINECO) through Grant no. FIS2011-23713 (F.G.), the European Research Council Advanced Grants  ``NOVGRAPHENE" (F.G.) and ``QUAGATUA" (M.L.), the Spanish Ministry of Science and Innovation (MINCIN) through the Grant ``TOQATA" (M.L.), the EU Integrated Project ``AQUTE" (M.L.), the US National Science Foundation through Grant DMR-1206916 (H.C.M.), the US Department of Energy, Office of Basic Energy Sciences, Division of Materials Sciences and Engineering, under contract DE-AC02-76SF00515 (H.C.M.), and the Italian Ministry of Education, University, and Research (MIUR) through the program 
``FIRB - Futuro in Ricerca 2010", Grant no. RBFR10M5BT (V.P. and M.P.).
\end{document}